# Multi-field Modelling of Cosserat solids

A. A. Vasiliev[1], A. E. Miroshnichenko[2, ✉]

[1]*Department of Mathematical Modelling, Tver State University,*
*Zhelyabov Street 33, 170000 Tver, Russia*
*e-mail: aleksey.vasiliev@tversu.ru*
[2]*Max-Planck-Institut für Physik komplexer Systeme,*
*Nythnitzer Strasse 38, D-01187 Dresden, Germany*
*e-mail: andreym@mpipks-dresden.mpg.de*
*Fax: +49 (351) 871-1999*

## SUMMARY

We constructed multi-field generalisations of the Cosserat continuum model on the basis of the square lattice model that takes into account rotational degree of freedom of microstructural elements. This approach allows us to model not only long but also short wavelength fields of displacements and rotations. The accuracy of the wave solutions is examined. The dispersion relations of plane harmonic waves for multi-field Cosserat continuum models are obtained and compared with those of the basic model.

**Key words:** Cosserat solid, Cosserat continuum model, short wavelength solutions

## 1. INTRODUCTION

Continuum models are effectively used for modelling of solids composed of a large number of microstructural elements and have a wide range of applications from nano-materials to space constructions. They are useful for prediction of physical-mechanical characteristics, obtaining of analytical solutions, constructing of efficient algorithms for numerical simulations, etc. But such models should be corrected in order to describe some essential microstructural effects. The development of more precise models is often based on analysis and rejection of some assumptions of the classical theories. Different classifications of the generalised models are given, for example, in /1, 2/.

Several types of generalised models are constructed by taking into account more degrees of freedom of structural elements than in classical models, e.g. /3-8/. In Cosserat and micropolar models not only space displacements but also independent rotations of the elements are taken into consideration and field of rotations in addition to the field of displacements is introduced. Structural models are useful for development and applications of field theories. The discrete models of materials with beam-microstructure and grid frameworks, corresponding micropolar models and methods for their obtaining are presented, for example, in /9-12/. The complex-valued Cosserat continuum model was constructed for vibrating grid frameworks in /5/. The variants of lattice models and corresponding Cosserat continuum models were obtained in /13, 14/. Via analysis of the dispersion curves, it has been demonstrated that the Cosserat continuum model gives accurate results for deformation fields with wavelengths larger than approximately six times micro-structural cell distance but produces unrealistic results for short wavelengths. In the present article, we construct generalisations of the Cosserat continuum model applicable not only for long but also for short wavelengths oscillating fields. As a rule, single cell of periodicity and one corresponding vector field of generalised displacements are used for obtaining of continuum models. Rejection of this not self-evident assumption leads to a new class of models. We construct $N$-field models on the basis of macrocell /15/, which is consist of $N$ simple cells. By using of $N$ vector fields we describe the deformations of the system. The macrocell method, the multi-field approach and the theory of coupled multi-continua, their applications to other tasks and models are presented in /15-21/. Another approach to construction more adequate models is to take into account a long range interaction (non-locality), for example. Higher-order gradient continuum models of granular materials and non-local Cosserat model of heterogeneous materials are presented in /22-24/.

## 2. DISCRETE MODEL

We consider a 9-cell lattice model (Fig. 1a), /13, 14/. Each particle possesses three degree of freedom - the displacements $u_k, v_k$ and rotations $\varphi_k$. The total potential energy is

$$E_{pot}^{(n,m)} = E_n^{(n,m)} + E_s^{(n,m)} + E_r^{(n,m)},$$

where

$$E_n^{(n,m)} = \frac{1}{2}K_n^{(n,m)}\left[(u_m - u_n)\cos\theta - (v_m - v_n)\sin\theta\right]^2,$$

$$E_s^{(n,m)} = \frac{1}{2}K_s^{(n,m)}\left[(u_m - u_n)\sin\theta + (v_m - v_n)\cos\theta + \frac{1}{2}d_{(n,m)}(\varphi_m + \varphi_n)\right]^2,$$

$$E_r^{(n,m)} = \frac{1}{2}G_r^{(n,m)}(\varphi_m - \varphi_n)^2.$$

Here, $\theta$ is the angle between the $x$-axis and the direction from the centre of mass of the particle $n$ to the centre of mass of the particle $m$; $K_n^{(n,m)}, K_s^{(n,m)}$ characterise the stiffnesses of the connection between particles in the longitudinal and shear directions; $G_r^{(n,m)}$ specifies the resistance to rotation. The elastic constants for axial connections are chosen equal to $K_n$, $K_s$, $G_r$ and the parameter $d_{(n,m)}$ equals the distance between the centres of particles $d$. We assume that $K_s^d = 0, G_r^d = 0$ and $K_n^d \neq 0$ for diagonal connections.

The kinetic energy of the particles is defined by

$$E_{kin}^n = \frac{1}{2}m\dot{u}_n^2 + \frac{1}{2}m\dot{v}_n^2 + \frac{1}{2}I\dot{\varphi}_n^2,$$

where $m$ is the mass and $I$ is the moment of inertia of the cell.

In order to obtain Lagrange's equations of $n$th particle, Lagrangian is constructed

$$L_n = \sum_m \left[ E_{kin}^n - E_{pot}^{(n,m)} \right],$$

where the summation is taken over all particles $m$, which contact with the particle $n$.

## 3 TWO-FIELD MODEL. THE CASE OF THE FACE-CENTRED SQUARE LATTICE

In order to derive a two-field model we consider a macrocell containing two elements (Fig. 1b) instead of the simple cell (Fig. 1a) that was used to construct the continuum model in /13, 14/ and employ two vector fields to describe the deformations of the system. In other words, the lattice is represented as a complex lattice or as two interpenetrating square lattices. Instead of real heterogeneity /6-8/, artificial heterogeneity is introduced. Different indexes 0 and 1 are used to mark elements although all elements are identical.

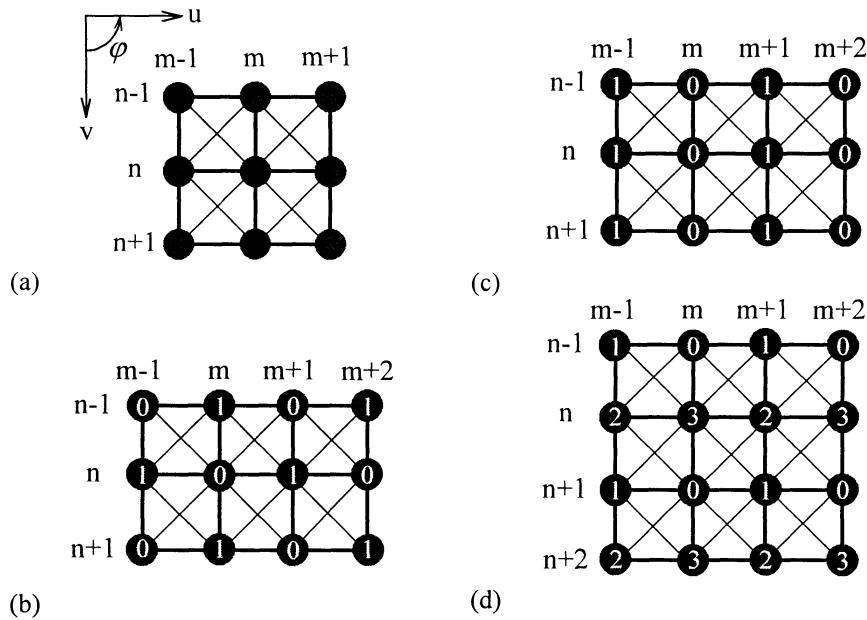

Fig. 1 a-d: 9-cell square lattice model. Different representations of square lattice as interpenetrating lattices

## 3.1 Equations and dispersion relations of discrete model

By using Lagrange's equations, we derive six equations of motion in the form

$$m\ddot{u}_r^{m,n} = K_n \left[ \Delta_{xx} u_j^{m,n} + 2u_j^{m,n} - 2u_r^{m,n} \right] + \frac{1}{2} K_n^d \left[ \Delta u_r^{m,n} + \Delta_{xy} v_r^{m,n} \right]$$

$$+ K_s \left[ \Delta_{yy} u_j^{m,n} + 2u_j^{m,n} - 2u_r^{m,n} - \frac{1}{2} d \Delta_y \varphi_j^{m,n} \right],$$

$$m\ddot{v}_r^{m,n} = K_n \left[ \Delta_{yy} v_j^{m,n} + 2v_j^{m,n} - 2v_r^{m,n} \right] + \frac{1}{2} K_n^d \left[ \Delta_{xy} u_r^{m,n} + \Delta v_r^{m,n} \right] \qquad (1)$$

$$+ K_s \left[ \Delta_{xx} v_j^{m,n} + 2v_j^{m,n} - 2v_r^{m,n} + \frac{1}{2} d \Delta_x \varphi_j^{m,n} \right],$$

$$I\ddot{\varphi}_r^{m,n} = G_r \left[ \Delta_{xx} \varphi_j^{m,n} + \Delta_{yy} \varphi_j^{m,n} + 4\varphi_j^{m,n} - 4\varphi_r^{m,n} \right]$$

$$+ \frac{1}{2} dK_s \left[ \Delta_y u_j^{m,n} - \Delta_x v_j^{m,n} - \frac{1}{2} d \left( \Delta_{xx} \varphi_j^{m,n} + \Delta_{yy} \varphi_j^{m,n} + 4\varphi_j^{m,n} + 4\varphi_r^{m,n} \right) \right],$$

where $r = 0, j = 1$ for the first three equations, $r = 1, j = 0$ and the index $m$ should be replaced by $m + 1$ for the last three equations; $u_k^{m,n}, v_k^{m,n}, \varphi_k^{m,n}$ define the displacements and rotation of the particle with coordinates $(md, nd)$ of the $k$th lattice. The following notations for finite differences have been used

$$\Delta_x w^{m,n} = w^{m+1,n} - w^{m-1,n}, \Delta_{xx} w^{m,n} = w^{m+1,n} - 2w^{m,n} + w^{m-1,n},$$

$$\Delta_y w^{m,n} = w^{m,n+1} - w^{m,n-1}, \Delta_{yy} w^{m,n} = w^{m,n+1} - 2w^{m,n} + w^{m,n-1},$$

$$\Delta w^{m,n} = w^{m+1,n+1} + w^{m+1,n-1} + w^{m-1,n+1} + w^{m-1,n-1} - 4w^{m,n},$$

$$\Delta_{xy} w^{m,n} = w^{m+1,n+1} - w^{m+1,n-1} - w^{m-1,n+1} + w^{m-1,n-1}.$$

We are looking for a solution of the system of the equations (1) in the form

$$u_k^{m,n} = \tilde{u}_k \exp\left[ i\left( \omega t - mK_x - nK_y \right) \right],$$

$$v_k^{m,n} = \tilde{v}_k \exp\left[ i\left( \omega t - mK_x - nK_y \right) \right], \qquad (2)$$

$$\varphi_k^{m,n} = \tilde{\varphi}_k \exp\left[ i\left( \omega t - mK_x - nK_y \right) \right],$$

where $K_x = k_x d$, $K_y = k_y d$; $k_x$ and $k_y$ are the wave numbers; $\tilde{u}_k, \tilde{v}_k, \tilde{\varphi}_k$ are the wave amplitudes, $k = 0, 1$; $\omega$ is the angular frequency.

By substituting the expressions (2) into Eqs. (1) we obtain the system of six homogeneous algebraic equations. This system can be split into two systems of three linear equations

$$\left[a_{k,1} + m\omega_k^2\right]\tilde{U}_k + a_{k,2}\tilde{V}_k + a_{k,3}i\tilde{\Phi}_k = 0,$$

$$a_{k,2}\tilde{U}_k + \left[a_{k,4} + m\omega_k^2\right]\tilde{V}_k + a_{k,5}i\tilde{\Phi}_k = 0, \tag{3}$$

$$a_{k,3}\tilde{U}_k + a_{k,5}\tilde{V}_k + \left[a_{k,6} + I\omega_k^2\right]i\tilde{\Phi}_k = 0,$$

where new variables are defined by the relations

$$\tilde{U}_k = \frac{1}{2}\left[\tilde{u}_1 + (-1)^k \tilde{u}_0\right], \tilde{V}_k = \frac{1}{2}\left[\tilde{v}_1 + (-1)^k \tilde{v}_0\right], \tilde{\Phi}_k = \frac{1}{2}\left[\tilde{\varphi}_1 + (-1)^k \tilde{\varphi}_0\right], k = 0,1. \tag{4}$$

The coefficients for $k = 0$ read

$$a_{0,1} = 2(\cos K_x - 1)K_n + 2(\cos K_x \cos K_y - 1)K_n^d + 2(\cos K_y - 1)K_s,$$

$$a_{0,2} = -2K_n^d \sin K_x \sin K_y, a_{0,3} = K_s d \sin K_y, a_{0,5} = -dK_s \sin K_x, \tag{5}$$

$$a_{0,4} = 2(\cos K_y - 1)K_n + 2(\cos K_x \cos K_y - 1)K_n^d + 2(\cos K_x - 1)K_s,$$

$$a_{0,6} = 2(\cos K_x + \cos K_y - 2)G_r + (-\cos K_x - \cos K_y - 2)d^2 K_s / 2.$$

The coefficients for $k = 1$ read

$$a_{1,1} = 2(-\cos K_x - 1)K_n + 2(\cos K_x \cos K_y - 1)K_n^d + 2(-\cos K_y - 1)K_s,$$

$$a_{1,2} = -2K_n^d \sin K_x \sin K_y, a_{1,3} = -dK_s \sin K_y, a_{1,5} = dK_s \sin K_x, \tag{6}$$

$$a_{1,4} = 2(-\cos K_y - 1)K_n + 2(\cos K_x \cos K_y - 1)K_n^d + 2(-\cos K_x - 1)K_s,$$

$$a_{1,6} = 2(-\cos K_x - \cos K_y - 2)G_r + (\cos K_x + \cos K_y - 2)d^2 K_s / 2$$

The coefficients $a_{0,m}$ coincide with the corresponding coefficients $a_{S,m}$ for simple cells that were obtained in /13, 14/, i.e. $a_{0,m} \equiv a_{S,m}$. Note that the coefficients $a_{1,m}$ can be derived by means of replacements $K_x \rightarrow K_x - \pi$, $K_y \rightarrow K_y - \pi$ in $a_{S,m}$.

Three dispersion surfaces of compression, shear and micro-rotational waves for a simple cell (Fig. 1a) are defined for $0 \leq K_x \leq \pi$, $0 \leq K_y \leq \pi$. Six surfaces for the discrete model in the case under consideration are defined for $K_x + K_y \leq \pi$, $K_x \geq 0$, $K_y \geq 0$. Three of them for coefficients $a_{0,m}$ coincide with dispersion surfaces for simple cells on this domain. Three other surfaces for $a_{1,m}$ can be obtained by reflections of the surfaces for simple cells defined in the domain $K_x + K_y \geq \pi$, $K_x \leq \pi$, $K_y \leq \pi$. The axis of reflection is defined by $K_x = \pi/2$, $K_y = \pi/2$.

### 3.2 Two-field equations

We introduce two vector functions $\{u_k(x,y,t), v_k(x,y,t), \varphi_k(x,y,t)\}$ in order to describe deformations of the corresponding lattices $k = 0, 1$. Replacements $w_k^{(m\pm 1, n\pm 1)} \to w_k(x\pm d, y\pm d)$ in Eqs. (1) lead to six equations, which are functional-difference with respect to space variables. Using the Taylor series expansions in the point where the equations are defined one can obtain an exact differential model. Taking into account only terms up to second order, we come to the approximating continuum model. Its equations can be obtained formally by means of replacements of finite differences in Eqs. (1) with derivatives

$$\Delta_x w_r^{m,n} \to 2d w_{r,x}, \Delta_{xx} w_r^{m,n} \to d^2 w_{r,xx}, \Delta_y w_r^{m,n} \to 2d w_{r,y}, \Delta_{yy} w_r^{m,n} \to d^2 w_{r,yy},$$

$$\Delta w_r^{m,n} \to 2d^2 \Delta w_r \equiv 2d^2 \left(w_{r,xx} + w_{r,yy}\right), \Delta_{xy} w_r^{m,n} \to 4d^2 w_{r,xy}.$$

The model consists of six coupled equations for interacting interpenetrating fields

$$mu_{r,tt} = K_n \left[ d^2 u_{j,xx} + 2u_j - 2u_r \right] + K_n^d d^2 \left[ \Delta u_r + 2v_{r,xy} \right]$$

$$+ K_s \left[ d^2 u_{j,yy} + 2u_j - 2u_r - d^2 \varphi_{j,y} \right],$$

$$mv_{r,tt} = K_n \left[ d^2 v_{j,yy} + 2v_j - 2v_r \right] + K_n^d d^2 \left[ 2u_{r,xy} + \Delta v_r \right] + K_s \left[ d^2 v_{j,xx} + 2v_j - 2v_r + d^2 \varphi_{j,x} \right], \tag{7}$$

$$I\varphi_{r,tt} = G_r \left[ d^2 \Delta \varphi_j + 4\varphi_j - 4\varphi_r \right] + K_s d^2 \left[ u_{j,y} - v_{j,x} - \frac{1}{4}\left( d^2 \Delta \varphi_j + 4\varphi_j + 4\varphi_r \right) \right].$$

where $r = 0$, $j = 1$ for the first three equations and $r = 1, j = 0$ for the other three ones.

Adding and subtracting the equations for indexes $(r = 0, j = 1)$ and $(r = 1, j = 0)$ and by introducing new variables

$$U_k = \frac{1}{2}\left[ u_1 + (-1)^k u_0 \right], V_k = \frac{1}{2}\left[ v_1 + (-1)^k v_0 \right], \Phi_k = \frac{1}{2}\left[ \varphi_1 + (-1)^k \varphi_0 \right], k = 0,1,$$

the system of six equations (7) can be split into two systems.

The equations for $k = 0$ have the form

$$mU_{0,tt} = K_n d^2 U_{0,xx} + K_n^d d^2 \left[ \Delta U_0 + 2V_{0,xy} \right] + K_s d^2 \left[ U_{0,yy} - \Phi_{0,y} \right],$$

$$mV_{0,tt} = K_n d^2 V_{0,yy} + K_n^d d^2 \left[ 2U_{0,xy} + \Delta V_0 \right] + K_s d^2 \left[ V_{0,xx} + \Phi_{0,x} \right], \tag{8}$$

$$I\Phi_{0,tt} = G_r d^2 \Delta \Phi_0 + K_s d^2 \left[ U_{0,y} - V_{0,x} - d^2 \Delta \Phi_0 / 4 - 2\Phi_0 \right].$$

These equations were obtained on the basis of simple cell in /13/. We retain the term $K_s d^4 \Delta \Phi_0/4$ in the third equation in order to have the second order approximation. The comparison of these equations and the equations of the Cosserat continuum model was made in /13/. We will not consider these problems of the one-field theories here.

For $k = 1$ we have new equations

$$mU_{1,tt} = K_n\left[-d^2U_{1,xx} - 4U_1\right] + K_n^d d^2\left[\Delta U_1 + 2V_{1,xy}\right] - K_s\left[d^2U_{1,yy} + 4U_1 - d^2\Phi_{1,y}\right],$$

$$mV_{1,tt} = K_n\left[-d^2V_{1,yy} - 4V_1\right] + K_n^d d^2\left[2U_{1,xy} + \Delta V_1\right] - K_s\left[d^2V_{1,xx} + 4V_1 + d^2\Phi_{1,x}\right], \quad (9)$$

$$I\Phi_{1,tt} = G_r\left[-d^2\Delta\Phi_1 - 8\Phi_1\right] + K_s d^2\left[-U_{1,y} + V_{1,x} + d^2\Delta\Phi_1/4\right].$$

In order to examine the model we consider plane wave solutions

$$U_k(x,y,t) = \tilde{U}_k \exp\left[i(\omega t - k_x x - k_y y)\right],$$

$$V_k(x,y,t) = \tilde{V}_k \exp\left[i(\omega t - k_x x - k_y y)\right], \quad (10)$$

$$\Phi_k(x,y,t) = \tilde{\Phi}_k \exp\left[i(\omega t - k_x x - k_y y)\right].$$

Substitution of (10) into Eqs. (9) leads to system (3) with coefficients

$$c_{0,1} = -K_x^2 K_n + \left(-K_x^2 - K_y^2\right)K_n^d - K_y^2 K_s, c_{0,2} = -2K_n^d K_x K_y,$$

$$c_{0,3} = dK_s K_y, c_{0,4} = -K_y^2 K_n + \left(-K_x^2 - K_y^2\right)K_n^d - K_x^2 K_s, \quad (11)$$

$$c_{0,5} = -dK_s K_x, c_{0,6} = \left(-K_x^2 - K_y^2\right)G_r + \left(K_x^2 + K_y^2 - 8\right)K_s d^2/4.$$

The coefficients $c_{0,m}$ are the Taylor series expansions of the coefficients $a_{0,m}$ for discrete model up to the second-order approximation around the point $k_x = 0$, $k_y = 0$.

For $k = 1$, the coefficients have the form

$$c_{1,1} = \left(K_x^2 - 4\right)K_n + \left(-K_x^2 - K_y^2\right)K_n^d + \left(K_y^2 - 4\right)K_s, c_{1,2} = -2K_n^d K_x K_y,$$

$$c_{1,3} = -dK_s K_y, c_{1,4} = \left(K_y^2 - 4\right)K_n + \left(K_x^2 - K_y^2\right)K_n^d + \left(K_x^2 - 4\right)K_s, \quad (12)$$

$$c_{1,5} = dK_s K_x, c_{1,6} = \left(K_x^2 + K_y^2 - 8\right)G_r + \left(-K_x^2 - K_y^2\right)K_s d^2/4.$$

The coefficients $c_{1,m}$ are the Taylor series expansions up to the second-order approximation of $a_{1,m}$ around the point $k_x = 0$, $k_y = 0$ and, at the same time, of coefficients $a_{S,m}$, Eqs. (5), for dispersion surfaces of simple square lattice around the point $k_x d = \pi$, $k_y d = \pi$. The relationship between $a_{S,\,m}$ and $a_{1,\,m}$ was established in Sec. 3.1.

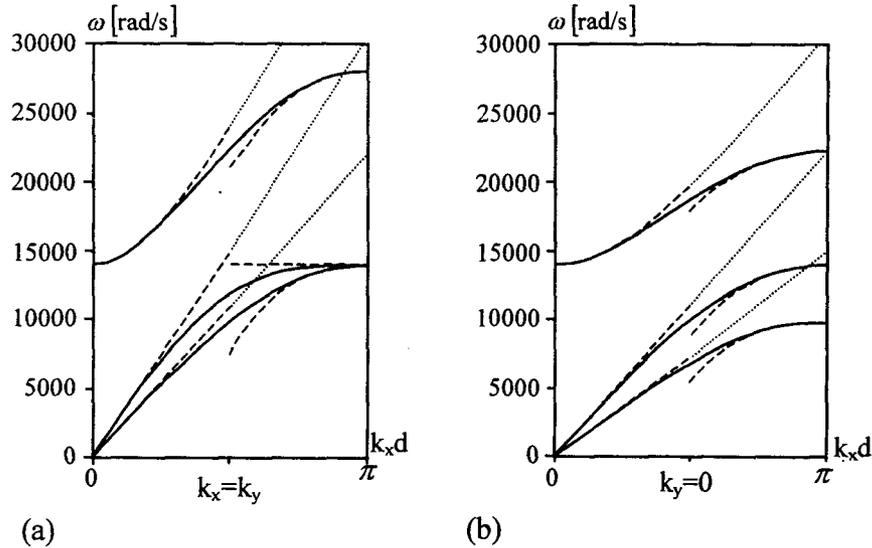

**Fig. 2 a,b:** Dispersion curves of shear, compression, micro-rotational waves for the lattice (solid line), Cosserat continuum model (dotted line), two-field models (dash line).

To illustrate the results of comparison, we present the section $K_x = K_y$ of the dispersion surfaces in Fig. 2a. The parameters are chosen the same as in /13, 14/ for granular medium: $\lambda = 55.5$ MPa, $\mu = 83.3$ MPa, $d = 0.05$ m, $G_r / 2d = 69.5$ kN, $K_n = 2d\mu$, $K_n^d = \lambda d$, $K_s = G_r / d^2$, $\rho = 1800$ kg/m$^3$, $J = 0.5625$ kg/m, $m = \rho d^3$, $I = Jd^3$. Dispersion curves for the lattice, Cosserat continuum model and two-field model are drawn using solid, dotted and dash lines, respectively. The curves for one-field Cosserat model and corresponding curves of two-field model coincide for long wavelengths in the interval $0 < K_x < \pi/2$. The two-field model gives a very good approximation for short wavelengths in the area near $K_x = K_y = \pi$ and it is exact at this point, where the one-field Cosserat model produces a rather large error.

## 4 TWO-FIELD MODEL. THE CASE OF TWO-LAYER LATTICE

We consider a simple lattice as a two-layer lattice (Fig. 1c).

### 4.1 Equations and dispersion relations for discrete model

By using Lagrange's equations, six equations of motion can be obtained

$$m\ddot{u}_r^{m,n} = \frac{1}{2}K_n^d \left[ \Delta u_j^{m,n} + 4u_j^{m,n} - 4u_r^{m,n} + \Delta_{xy}v_j^{m,n} \right]$$

$$+ K_n \left[ \Delta_{xx}u_j^{m,n} + 2u_j^{m,n} - 2u_r^{m,n} \right] + K_s \left[ \Delta_{yy}u_r^{m,n} - d\Delta_y \varphi_r^{m,n}/2 \right],$$

$$m\ddot{v}_r^{m,n} = K_n \Delta_{yy} v_r^{m,n} + \frac{1}{2} K_n^d \left[ \Delta_{xy} u_j^{m,n} + \Delta v_j^{m,n} + 4v_j^{m,n} - 4v_r^{m,n} \right]$$

$$+ K_s \left[ \Delta_{xx} v_j^{m,n} + 2v_j^{m,n} - 2v_r^{m,n} + \frac{1}{2} d \Delta_x \varphi_j^{m,n} \right], \quad (13)$$

$$I\ddot{\varphi}_r^{m,n} = G_r \left[ \Delta_{yy} \varphi_r^{m,n} + \Delta_{xx} \varphi_j^{m,n} - 2\varphi_r^{m,n} + 2\varphi_j^{m,n} \right]$$

$$+ \frac{1}{2} dK_s \left[ \Delta_y u_r^{m,n} - \Delta_x v_j^{m,n} - \frac{1}{2} d \left( \Delta_{xx} \varphi_j^{m,n} + \Delta_{yy} \varphi_r^{m,n} + 2\varphi_j^{m,n} + 6\varphi_r^{m,n} \right) \right].$$

where $r = 0, j = 1$ for the first three equations and $r = 1, j = 0, m \to m + 1$ for the last three equations.

Substitution of (2) ($0 \leq K_x \leq \pi/2$, $0 \leq K_y \leq \pi$) into Eqs. (13) leads to the system of six algebraic equations for amplitudes, which can be split into two systems (3). The coefficients corresponding to $k = 0$ in (4) are the same as coefficients $a_{0,m}$. The coefficients corresponding to the subtraction are as follows:

$$a_{2,1} = 2(-\cos K_x - 1)K_n + 2(-\cos K_x \cos K_y - 1)K_n^d + 2(\cos K_y - 1)K_s,$$

$$a_{2,2} = 2K_n^d \sin K_x \sin K_y, a_{2,3} = dK_s \sin K_y, a_{2,5} = dK_s \sin K_x, \quad (14)$$

$$a_{2,4} = 2(\cos K_y - 1)K_n + 2(-\cos K_x \cos K_y - 1)K_n^d + 2(-\cos K_x - 1)K_s,$$

$$a_{2,6} = 2(-\cos K_x + \cos K_y - 2)G_r + (\cos K_x - \cos K_y - 2)d^2 K_s/2.$$

These coefficients could be obtained from $a_{S,m}$, Eqs. (5), by the replacement $K_x \to K_x - \pi$.

### 4.2 Two-field equations

In order to construct the two-field model we use two vector functions. Six differential equations can be derived from Eqs. (13) by using the method described in Sec. 3.2. In the linear case the system of six coupled equations can be split into two systems. The first system, corresponding to addition, is system (8). Therefore, the two-field model contains the equations of the Cosserat theory. New equations, corresponding to subtraction, have the form

$$mU_{2,tt} = -K_n \left[ d^2 U_{2,xx} + 4U_2 \right] - K_n^d \left[ d^2 \Delta U_2 + 4U_2 + 2d^2 V_{2,xy} \right] + K_s d^2 \left[ U_{2,yy} - \Phi_{2,y} \right],$$

$$mV_{2,tt} = K_n d^2 V_{2,yy} - K_n^d \left[ d^2 \Delta V_2 + 4V_2 + 2d^2 U_{2,xy} \right] - K_s \left[ d^2 V_{2,xx} + 4V_2 + d^2 \Phi_{2,x} \right],$$

$$I\Phi_{2,tt} = G_r \left[ -d^2 \Phi_{2,xx} + d^2 \Phi_{2,yy} - 4\Phi_2 \right]$$

$$+ K_s d^2 \left[ U_{2,y} + V_{2,x} + d^2 \left( \Phi_{2,xx} - \Phi_{2,yy} \right)/4 - \Phi_2 \right]. \quad (15)$$

The coefficients for dispersion relations are

$$c_{2,1} = \left(K_x^2 - 4\right)K_n + \left(K_x^2 + K_y^2 - 4\right)K_n^d - K_y^2 K_s, c_{2,2} = 2K_n^d K_x K_y,$$

$$c_{2,3} = dK_s K_y, c_{2,4} = -K_y^2 K_n + \left(K_x^2 + K_y^2 - 4\right)K_n^d + \left(K_x^2 - 4\right)K_s,$$

$$c_{2,5} = dK_s K_x, c_{2,6} = \left(K_x^2 - K_y^2 - 4\right)G_r + \left(-K_x^2 + K_y^2 - 4\right)K_s d^2 / 4. \tag{16}$$

The coefficients $c_{2,m}$ are the Taylor series expansions up to second-order approximation of the coefficients $a_{2,m}$ around the point $K_x = 0$, $K_y = 0$ and of the coefficients $a_{S,m}$ for simple square lattice around the point $k_x d = \pi$, $k_y d = 0$.

Therefore, the two-field model contains the Cosserat model and gives accurate results as for long wavelengths and for short wavelengths around the point $k_x d = \pi$, $k_y d = 0$ as well.

To illustrate this, we present the section $K_y = 0$ of the dispersion surfaces in Fig. 2b. The parameters and style of lines are the same as in Fig. 2a. The two-field model has the same accuracy for modelling of long wavelength fields in discrete system as the Cosserat continuum model. It can be also used in the domain of the short wavelengths near $K_x = \pi$, $K_y = 0$ and is exact at this point.

**Remark.** The equations (8), (18) and the coefficients for dispersion relations (11), (19) of the two-field model for the case of two-layer square lattice when the layers are parallel to the x-axis will be obtained in the next section.

## 5 FOUR-FIELD MODEL

In order to construct a four-field model we consider a macrocell containing four microstructural elements numbered by indexes from 0 to 3 (Fig. 1d).

By using Lagrange's equations we derive twelve equations of motion for particles of the macrocell. The dispersion relations can be obtained by the substitution of the Eqs. (2), $k = \overline{0,3}$, into these equations. The system of 12 algebraic equations can be decomposed into four systems for new variables

$$\tilde{U}_0 = \frac{1}{4}[\tilde{u}_0 + \tilde{u}_1 + \tilde{u}_2 + \tilde{u}_3], \ \tilde{U}_1 = \frac{1}{4}[-\tilde{u}_0 + \tilde{u}_1 - \tilde{u}_2 + \tilde{u}_3],$$

$$\tilde{U}_2 = \frac{1}{4}[-\tilde{u}_0 + \tilde{u}_1 + \tilde{u}_2 - \tilde{u}_3], \ \tilde{U}_3 = \frac{1}{4}[\tilde{u}_0 + \tilde{u}_1 - \tilde{u}_2 - \tilde{u}_3], \tag{17}$$

and similar formulae for $\tilde{V}_k, \tilde{v}_k$ and $\tilde{\Phi}_k, \tilde{\varphi}_k$. The coefficients $a_{0,m}, a_{1,m}, a_{2,m}$ in (3) are defined by expressions (5), (6), (14), respectively. The coefficients $a_{3,m}$ can be obtained by using the replacement $K_y \to K_y - \pi$ in $a_{S,m}$, which are defined by (5).

Twelve dispersion surfaces are defined in the domain $G_0$ and consist of four groups. The first three surfaces for $k = 0$ coincide inside $G_0$ with dispersion surfaces for the model that was obtained on the basis of a simple cell. The dispersion surfaces for $k = 1$ are the reflections of surfaces for a simple cell defined inside the domain $G_1$. The axis of reflection is defined by $K_x = \pi/2$, $K_y = \pi/2$. The dispersion surfaces for $k = 2$ and $k = 3$ are reflections of the surfaces for simple cell that are defined inside the domains $G_2$ and $G_3$, respectively. The reflection planes are $K_x = \pi/2$, $K_y = \pi/2$. Square domains $G_k$ are denoted in Fig. 3.

We derive the system of twelve coupled equations of the four-field model by using four vector fields $\{u_k(x,y,t), v_k(x,y,t), \varphi_k(x,y,t)\}$ $k = \overline{0,3}$. This system can be decomposed into four systems for new variables $U_k$, $V_k$, $\Phi_k$, $k = \overline{0,3}$, which are defined by formulae (17). The equations for $k = 0, 1, 2$ are equations (8), (9), (15), respectively.

The new system for $k = 3$ has the form

$$mU_{3,tt} = K_n d^2 U_{3,xx} - K_n^d \left[ d^2 \Delta U_3 + 4U_3 + 2d^2 V_{3,xy} \right] - K_s \left[ d^2 U_{3,yy} + 4U_3 - d^2 \Phi_{3,y} \right],$$

$$mV_{3,tt} = K_n \left[ -d^2 V_{3,yy} - 4V_3 \right] - K_n^d \left[ d^2 \Delta V_3 + 4V_3 + 2d^2 U_{3,xy} \right] + K_s d^2 \left[ V_{3,xx} + \Phi_{3,x} \right],$$

$$I\Phi_{3,tt} = G_r \left[ d^2 \left( \Phi_{3,xx} - \Phi_{3,yy} \right) - 4\Phi_3 \right] - K_s d^2 \left[ U_{3,y} + V_{3,x} + \left( \Phi_{3,xx} - \Phi_{3,yy} \right) d^2 / 4 + \Phi_3 \right]. \tag{18}$$

The coefficients for dispersion relations are

$$c_{3,1} = -K_x^2 K_n + \left( K_x^2 + K_y^2 - 4 \right) K_n^d + \left( K_y^2 - 4 \right) K_s, \; c_{3,2} = 2K_n^d K_x K_y,$$

$$c_{3,3} = -dK_s K_y, c_{3,4} = \left( K_y^2 - 4 \right) K_n + \left( K_x^2 + K_y^2 - 4 \right) K_n^d - K_x^2 K_s, \tag{19}$$

$$c_{3,5} = -dK_s K_x, c_{3,6} = \left( -K_x^2 + K_y^2 - 4 \right) G_r + \left( K_x^2 - K_y^2 - 4 \right) K_s d^2 / 4.$$

These coefficients are Taylor series expansions of the coefficients $a_{3,m}$ up to second order approximation around the point $k_x d = 0$, $k_y d = 0$ and of the coefficients for simple lattice $a_{S,m}$ around the point $k_x d = 0$, $k_y d = \pi$.

Hence, the four-field model contains the equations of the Cosserat model and both two-field models obtained above. Additionally, it contains a new system (19) corresponding to the two-field model for the case when the layers are parallel to the $x$-axis.

The results are shown in Fig. 3. Parameters of the discrete system are chosen the same as for Fig. 2. The borderline of the dispersion surface of micro-rotational waves in discrete system is drawn above the border of the area $0 < K_x < \pi$, $0 < K_y < \pi$, where this surface is defined. Four surfaces for micro-rotational waves of the four-field model corresponding to (11), (12), (16), (19) are plotted inside the domains $G_0$, $G_1$, $G_2$, $G_3$, respectively. There is a good approximation near the corners of Brillouin zone. The solid line on the $0k_xk_y$ plane describes the domain near the corners where the ratio error for frequencies is less than 10%.

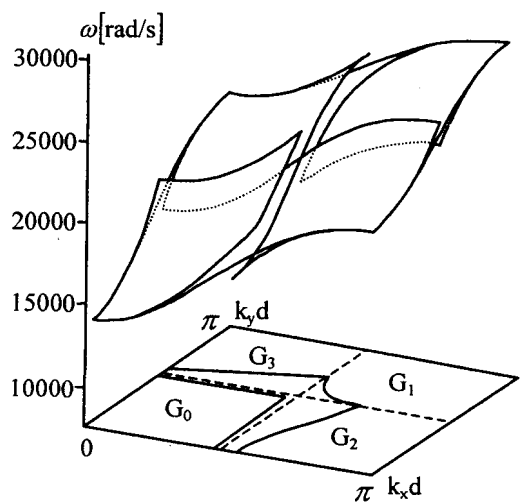

Fig. 3: The dispersion surfaces of the micro-rotational waves of the square lattice and four-field Cosserat model

## CONCLUSION

In this article we develop the multi-field approach for modelling of Cosserat solids, demonstrate technique and, as result, construct hierarchical set of multi-field Cosserat continuum models, which describe the dynamical properties with increasing accuracy. To create $N$-field models, we represent a Cosserat solid as $N$ interpenetrating ones and use $N$ vector fields of the same structure that is used in the Cosserat model to describe their deformations. Two kinds of two-field models were constructed. It was shown that both models contain the one-field Cosserat continuum model and the additional equations for short waves were extracted. The four-field model combines the conventional Cosserat continuum model and two-field models into one, possesses the important dynamical properties of these models and can be applied to modelling for long and short wavelength fields as well.